\documentclass[conference]{IEEEtran}
\IEEEoverridecommandlockouts
\usepackage{algorithm}
\usepackage{algorithmic}
\usepackage[mode=buildnew]{standalone}
\usepackage{amsmath,mathtools,nccmath}
\usepackage{siunitx}
\usepackage{cite}

\usepackage{dblfloatfix}
\usepackage{booktabs}
\usepackage{multirow}

\usepackage{enumitem}
\usepackage{mathtools}
\usepackage[utf8]{inputenc}
\usepackage{amssymb}
\usepackage{gensymb}

\usepackage[compact]{titlesec}
    \titlespacing{\section}{0pt}{2ex}{1ex}
    \titlespacing{\subsection}{0pt}{1ex}{0ex}
    \titlespacing{\subsubsection}{0pt}{0.5ex}{0ex}
\usepackage{float}
\usepackage{array}
\usepackage{algorithmic}
\usepackage{textcomp}
\usepackage{xcolor}
\usepackage{tabularx}
\def\BibTeX{{\rm B\kern-.05em{\sc i\kern-.025em b}\kern-.08em
    T\kern-.1667em\lower.7ex\hbox{E}\kern-.125emX}}

\begin{document}

\include{bibliography}

\title{Stability Enhancement of LCL-Type Grid-Following Inverters Using Capacitor Voltage Active Damping\\
}

\author{
\IEEEauthorblockN{Naser Souri,$^1$ \emph{Graduate Student Member, IEEE}, Ali Mehrizi-Sani,$^1$ \emph{Senior Member, IEEE},\\  Kambiz Tehrani,$^2$ \emph{Senior Member, IEEE}}
\IEEEauthorblockA{$^1$The Bradley Department of Electrical and Computer Engineering 
Virginia Tech, Blacksburg, VA 24061 \\
\IEEEauthorblockA{$^2$Department of Energy and Control, University of Normandy-ESIGELEC-IRSEEM, Rouen, France}
e-mails: \{nsouri,mehrizi\}@vt.edu, tehrani@esigelec.fr}

}

\maketitle

\begin{abstract}
An LCL filter offers superior attenuation for high-frequency harmonics for three-phase grid-following inverters compared to LC and L filters. However, it also introduces an inherent resonance peak, which can lead to power quality issues or even instability of the inverter control system. Active damping (AD) is widely employed to effectively mitigate this resonance. Capacitor voltage feedback (CVF) and capacitor current feedback (CCF) are effective AD methods for LCL resonance damping. CVF is preferred due to its lower sensor requirement compared to CCF. However, a derivative term appears in the active damping loop, which introduces high-frequency noise into the system. This paper proposes a noise-immune approach by replacing the derivative term with a discrete function suitable for digital implementation. The LCL resonance can be damped effectively, resulting in enhanced stability of the inverter control system. Simulation results verify the proposed effectiveness of the method with grid inductance variation and weak grid conditions \footnote{“© 20XX IEEE.  Personal use of this material is permitted.  Permission from IEEE must be obtained for all other uses, in any current or future media, including reprinting/republishing this material for advertising or promotional purposes, creating new collective works, for resale or redistribution to servers or lists, or reuse of any copyrighted component of this work in other works.”
}.

\end{abstract}


\section{Introduction}
Grid-following (GF) inverters play a crucial role in integrating renewable energy sources, including solar and wind power, into the electrical grid \cite{Choolabi, Farhangi, Ali}. GF inverters are required to have a filter at their terminals to effectively mitigate the harmonics caused by the PWM modulation effect. Among the filters, the most effective choice for harmonic attenuation and increasing power quality is a third-order LCL filter due to its steep attenuation slope of -40~dB/dec for high-frequency harmonics \cite{saleh}. 
Although this filter offers superior harmonic rejection, it may cause the grid current to oscillate or potentially introduce instability and damage to the inverter and other components in the system. The occurrence of oscillations or instability stems from resonance between the inverter and the grid, where the inverter's switching frequency coincides with the natural frequency of the filter. To address this issue, inverters are required to be equipped with advanced controllers to prevent such issues. Two prevalent techniques for resonance damping are:

\begin{itemize}
\item Passive damping: Passive damping, which is simple in design and implementation, involves incorporating additional components, such as resistors, into the LCL filter in series or parallel configurations, such as a resistor, to reduce the resonance.
\item Active damping: AD employs an extra path in the inverter control or utilizes a state feedback loop of the LCL current or voltage and generates an appropriate control signal to suppress the resonance.
\end{itemize}

Passive methods introduce losses in the system and reduce efficiency. Furthermore, they may not provide adequate damping at high frequencies \cite{Mehrdad}. In contrast, active damping methods effectively suppress resonance and minimize harmonic distortion. Numerous research studies explore active damping strategies employing state feedback on the LCL components. For instance, \cite{Hanif} employs a grid-side current with a low-pass filter, while \cite{Guan} uses the inverter-side current to mitigate the resonance in the system. References \cite{Dannehl}-\cite{Javier} utilize filters in the control path to provide damping; however, they deteriorate the performance or exhibit a strong dependency on the resonance frequency. Reference \cite{Kumar} proposes an accurate digital differentiator, albeit applicable primarily to low-frequency applications. As frequency increases, the magnitude deviation from a real differentiator becomes more pronounced. Reference \cite{Chenlei} employs capacitor current feedback with a proportional gain to suppress the resonance; however, this approach detrimentally affects the phase margin. Thus, \cite{Yuying} utilizes a high-pass filter (HPF) instead. However, the HPF technique introduces high-frequency harmonics into the control loop. Reference \cite{Azghandi} utilizes fractional-order active damping for the term s in the Laplace domain to diminish gain for noisy signals and enhance the damping ratio; nonetheless, this approach is relatively complex. References \cite{Rafael, SHAO2003157} employ a lead-lag compensator and wavelet transform, which serve as a differentiator. However, these methods exhibit effectiveness within a limited frequency range. Reference \cite{Mingming} utilizes a band-pass filter (BPF) equipped with a lead compensator as a differentiator. However, it features a relatively narrow band for differentiation. Reference \cite{Donghua} presents a method to differentiate employing a quadrature-second-order generalized integrator (Q-SOGI). By altering the cutoff frequency to stabilize the phase, this technique achieves a remarkable increase in magnitude. On the other hand, to reduce the number of current sensors \cite{Sirat}, capacitor voltage can be measured for both synchronization and AD instead. Therefore, further research is needed to be conducted.


This paper proposes a discrete transfer function as a differentiator in the AD loop to emulate the capacitor current. Following is the remainder of the paper:

Section II provides a system description for both the grid-connected inverter and the LCL resonance.
Section III delves into the available differentiators in the discrete domain and discusses the proposed differentiator. Section IV presents a comprehensive stability analysis of the closed-loop system equipped with CVF-AD. Section V discusses different cases to evaluate how well the technique works. In this section, a simulation is performed to validate the proposed method.


\section{LCL-Type Grid-Connected Inverter}
\begin{figure}[htbp]
\centerline{\includegraphics[width= 0.8\columnwidth ]{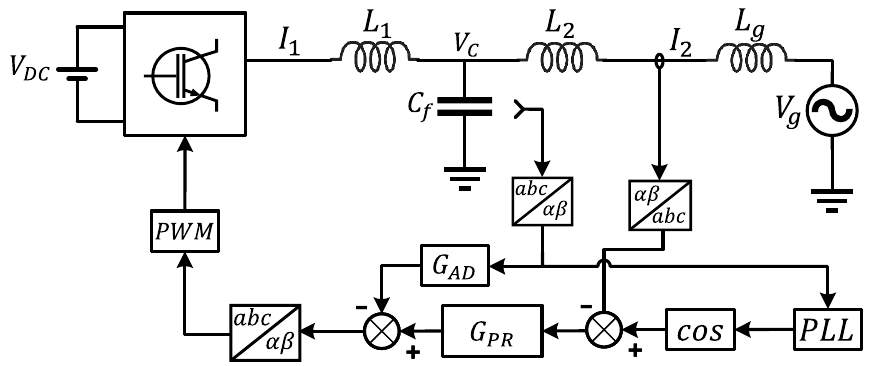}}
\vspace*{-0.1 cm}
\caption{Three-phase grid-following inverter block diagram equipped with AD.}
\label{fig:sys_block_diagram}
\end{figure}

\begin{figure}[htbp]
\centerline{\includegraphics[width= 0.9\columnwidth ]{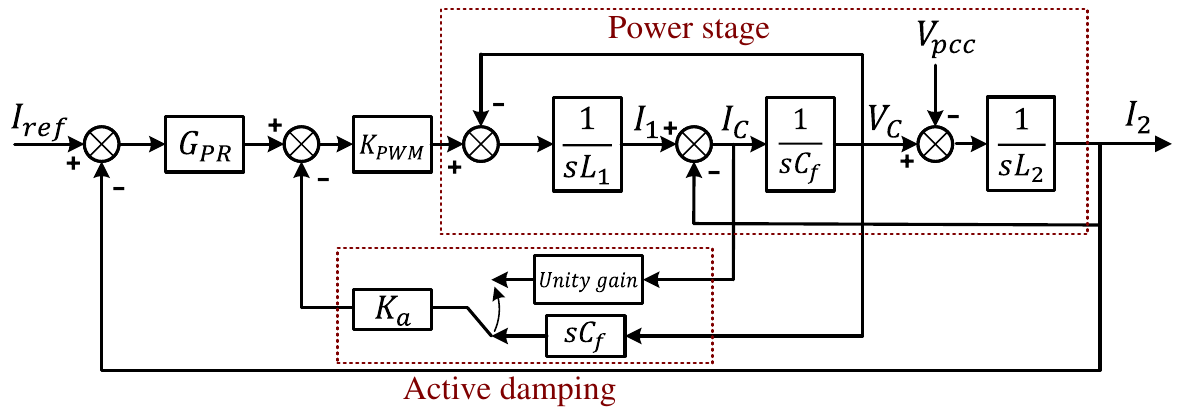}}
\vspace*{-0.2 cm}
\caption{Inverter control loop block diagram equipped with CVF-AD.}
\label{fig:control_block_diagram}
\end{figure}

An LCL filter connects a three-phase inverter to the grid, as depicted in Fig.~\ref{fig:sys_block_diagram}. The inverter topology employs a three-phase H-bridge converter. $I_1$ and $I_2$ are the currents from the grid and the inverter sides, respectively. $L_g$ is the grid inductance, which can vary in value. Fig.~\ref{fig:control_block_diagram} illustrates the block diagram for the inverter control. This control scheme encompasses the inverter control, the LCL filter, and the AD loop. $\alpha\beta$ coordination is employed for the inverter control loop. $I_{ref}$ is the reference current in $\alpha\beta$ coordination. $G_c$ is the proportional resonant controller, and its transfer function is:

\begin{equation}
G_c=K_p+\frac{K_rs}{s^2+\omega_c^2}
\end{equation}

$K_p$ denotes the controller gain, $K_r$ is the resonance gain, and $\omega_c$ is the fundamental frequency based on rad/s.
The LCL filter transfer function from $I_2$ to $V_{o}$ is calculated by:

\begin{equation}
G_{LCL}=\frac{I_2}{V_o}=\frac{1}{L_1(L_2+L_g)C_f}\frac{1}{s(s^2+\omega_r^2)}
\end{equation}

Where it has a resonance at $\omega_r$, which can be calculated from:

\begin{equation}
\omega_r=\sqrt{\frac{L_1+L_2}{L_1(L_2+L_g)C_f}}
\end{equation}

\begin{figure}[htbp]
\centerline{\includegraphics[width= 0.85\columnwidth ]{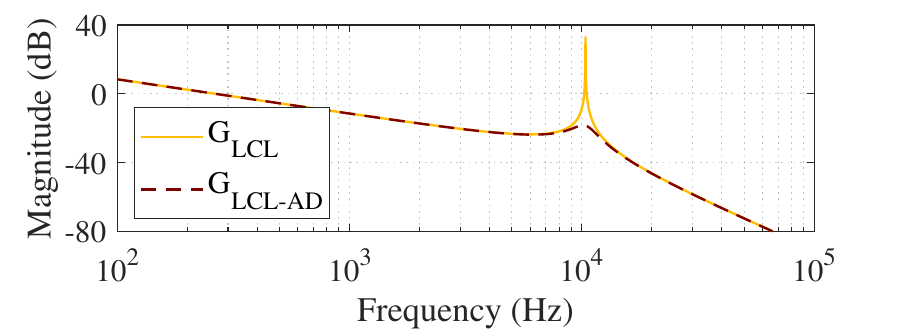}}
\vspace*{-0.2 cm}
\caption{Frequency response of an LCL compared with a damped LCL.}
\label{fig:LCL_freq_responce}
\end{figure}

Fig.~\ref{fig:LCL_freq_responce} depicts the LCL filter's frequency response. The resonance manifests as a peak gain in this figure. By employing the inductor or capacitor currents or voltages as a feedback state in the AD loop, a damping term emerges in the LCL filter transfer function; hence, AD effectively mitigates the peak resonance.

\begin{equation}
G_{LCL-AD}=\frac{K_f}{s(s^2+\frac{k_a}{L_1}s+\omega_r^2)}
\end{equation}

Where $K_f=\frac{1}{L_1{L'_2 C_f}}$, and  ${L'}_2=L_2+L_g$. \\ \\
$K_a$ is the damping gain in the active damping feedback loop. 

\section{Discrete Differentiators and Filters}

\subsection{Backward Euler Differentiator}
The backward Euler method is a one-step method, meaning that it only requires the current value of the ODE to compute the next value. The backward transfer function is:

\begin{equation}
G_{Back}\left(z\right)=\frac{z-1}{T_sz}
\end{equation}

\subsection{Forward Euler Differentiator} 
Forward Euler is another method for the discretization of a function. The forward method implementation is easy, though it adds unstable poles to the system while discretizing a function. As a result, this approach is not applicable.

\subsection{Tustin Differentiator}
The Tustin method, also recognized as the bilinear transform, is another technique for discretizing. This method is widely employed due to its ability to maintain system stability and offer a more accurate approximation than other methods. In the z-domain, Tustin's transfer function is:

\begin{equation}
G_{Tustin}\left(z\right)=\frac{2}{T_s}\frac{z-1}{z+1}
\end{equation}

\begin{figure}[!t]
\includegraphics[width= 0.96\columnwidth]{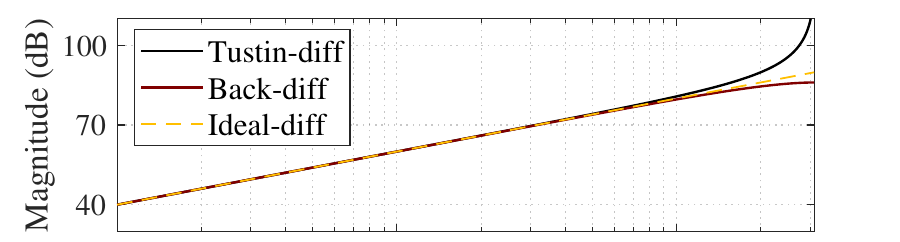}
\includegraphics[width= 0.96\columnwidth]{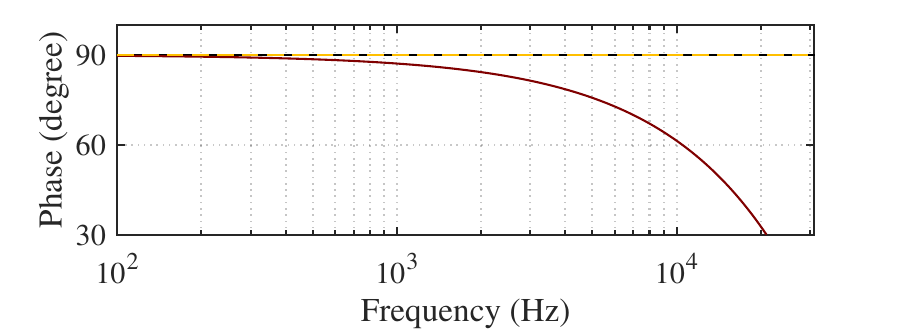}
\vspace*{-0.4 cm}
\caption{Comparison of different differentiators with a real derivative.} \label{fig:discrete_Diff_comparision}
\end{figure}

Fig.~\ref{fig:discrete_Diff_comparision} contrasts the discrete backward and Tustin differentiator's frequency responses with a real differentiator. The frequency response for the Tustin demonstrates that the phase remains $90$~degrees in all frequencies, though it has a significant amplification of the gain near the Nyquist frequency, which causes noise amplification and instability.
The backward method, on the other hand, presents superior gain characteristics. Yet, its phase response worsens as frequency increases. Additionally, the forward method renders the system unstable. Consequently, these methods are not suitable replacements for a genuine differentiator. To enhance their performance, additional filters or compensators, such as a lead compensator, which will be discussed in the following section, must be incorporated into the differentiators.

\subsection{Lead Compensator} 
A lead compensator is a type of filter that can alter the gain and phase of a system at a specific frequency range. It is frequently employed to enhance the performance of control systems by increasing the bandwidth of the system and making it more responsive. A discrete lead compensator in the z-domain is:

\begin{equation}
G_{Lead}=P_z\frac{z}{z+P_z}
\end{equation}

\begin{figure}[t]
\includegraphics[width= 0.95\columnwidth]{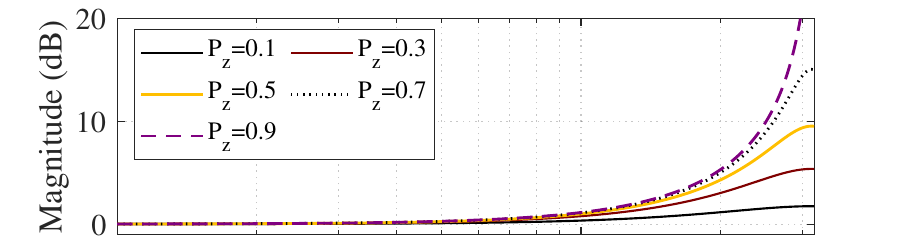}
\includegraphics[width= 0.95\columnwidth]{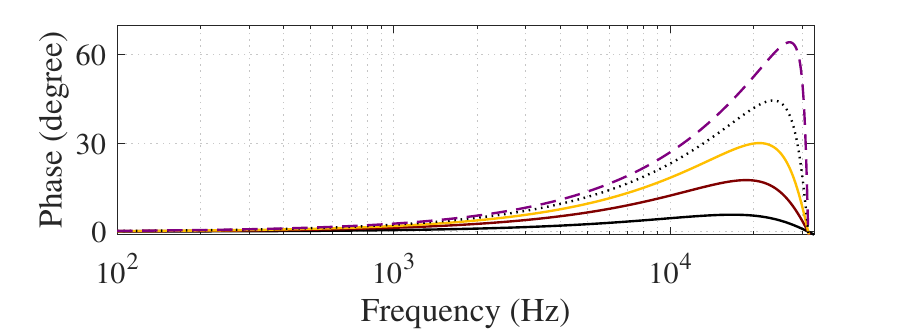}
\vspace*{-0.4 cm}
\caption { A lead compensator's frequency response with various designs.}
\vspace*{-0.2 cm}

\label{fig:Lead_compensator_design}
\end{figure}

Where Pz is the pole of the discrete lead transfer function. Based on the value of Pz, the gain and the phase compensation of the lead transfer function change. Fig.~\ref{fig:Lead_compensator_design} shows these changes based on $P_z$. Then, according to the required phase compensation, $P_z$ is selected.


\subsection{Notch Filter} 
A notch filter is a type of filter that selectively attenuates specific frequencies within a narrow range while allowing all other frequencies to pass through unimpeded. The transfer function of a second-order notch filter is as follows, which is in the z-domain:

\begin{equation}
G_{DNF}=\frac{(m+1)(z+1)(2z-1)}{\left(2m+2\right)z^2+z-1}
\end{equation}

Where $m$ determines the stop-band range.

\subsection{The Proposed Differentiator}
As mentioned in the previous sections, none of the discrete differentiators are sufficient to mimic a real differentiator alone. A lead compensator can improve their performance. By adding a lead compensator to the backward method, the phase lag decreases:

\begin{equation}
G_{Back-Lead}=\left(\frac{z-1}{T_sz}\right)\left(P_z\frac{z}{z-P_z}\right)=\frac{P_z}{T_s}\frac{z-1}{z-P_z}
\end{equation}

By increasing $P_z$, it imparts phase to the differentiator, and the phase error decreases; however, the gain near the Nyquist grows significantly. This work endeavors to further elevate $P_z$ to compensate for phase lag and then incorporate a notch filter to resolve the gain amplification issue. The notch filter, therefore, can be designed at the Nyquist frequency to damp the gain amplification caused by increasing $P_z$. Consequently, the proposed differentiator transfer function is derived:


\begin{equation}
G_{diff}=\frac{P_z}{T_s}\frac{(m+1)(z^2+1)(2z-1)}{(\left(2m+2\right)z^2+z-1)(z-P_z)}
\label{diff_pro}
\end{equation}

\begin{figure}[t]
\includegraphics[width=0.95\linewidth] 
{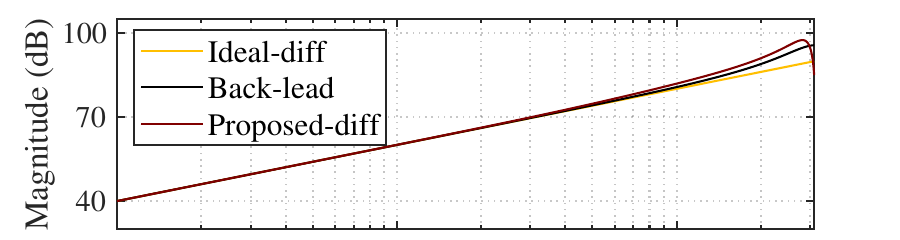}   
\includegraphics[width=0.95\linewidth]{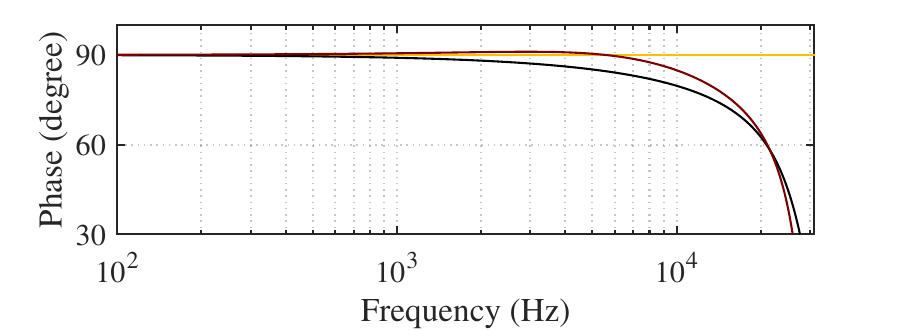}

\vspace*{-0.4 cm}
\caption{Proposed differentiator in comparison with backward-lead and ideal.} \label{fig:proposed_discrete_diff}
\end{figure}

The frequency response of (\ref{diff_pro}) in Fig.~\ref{fig:proposed_discrete_diff} depicts how the proposed method effectively emulates a real differentiator within the intended area of differentiation. Compared to the backward plus lead compensator that \cite{Donghua} proposes, it exhibits superior accuracy. The gain of the backward-Tustin-notch is slightly higher at higher frequencies, but it does not make very much of a difference, and it matches a real differentiator at the lower frequencies. The zone of differentiating in this paper is around $f_{res}=2.8$~kHz. $G_{AD}$ is the AD gain presented in the control block diagram of Fig.~\ref{fig:control_block_diagram}. $G_{AD}$ is calculated from (\ref{AD_gain}).

\begin{equation}
G_{AD}= C_f \times G_{diff}
\label{AD_gain}
\end{equation}

\section{Stability Analysis}

A stability analysis is conducted regarding the variation of active damping gain to see the pole-zeros placement. The stability analysis is performed for the control block diagram demonstrated in Fig.~\ref{fig:control_block_diagram}. The Laplace domain controller's open-loop transfer function is:

\begin{equation}
G_{OL}=(K_p+\frac{K_is}{s^2+\omega_c^2})(\frac{K_{PWM}K_f}{s(s^2+\frac{{K_{PWM}K}_a}{L_1}s+\omega_r^2)})
\end{equation}

\begin{figure}[t]
\centerline{\includegraphics[width= 0.65\columnwidth ]{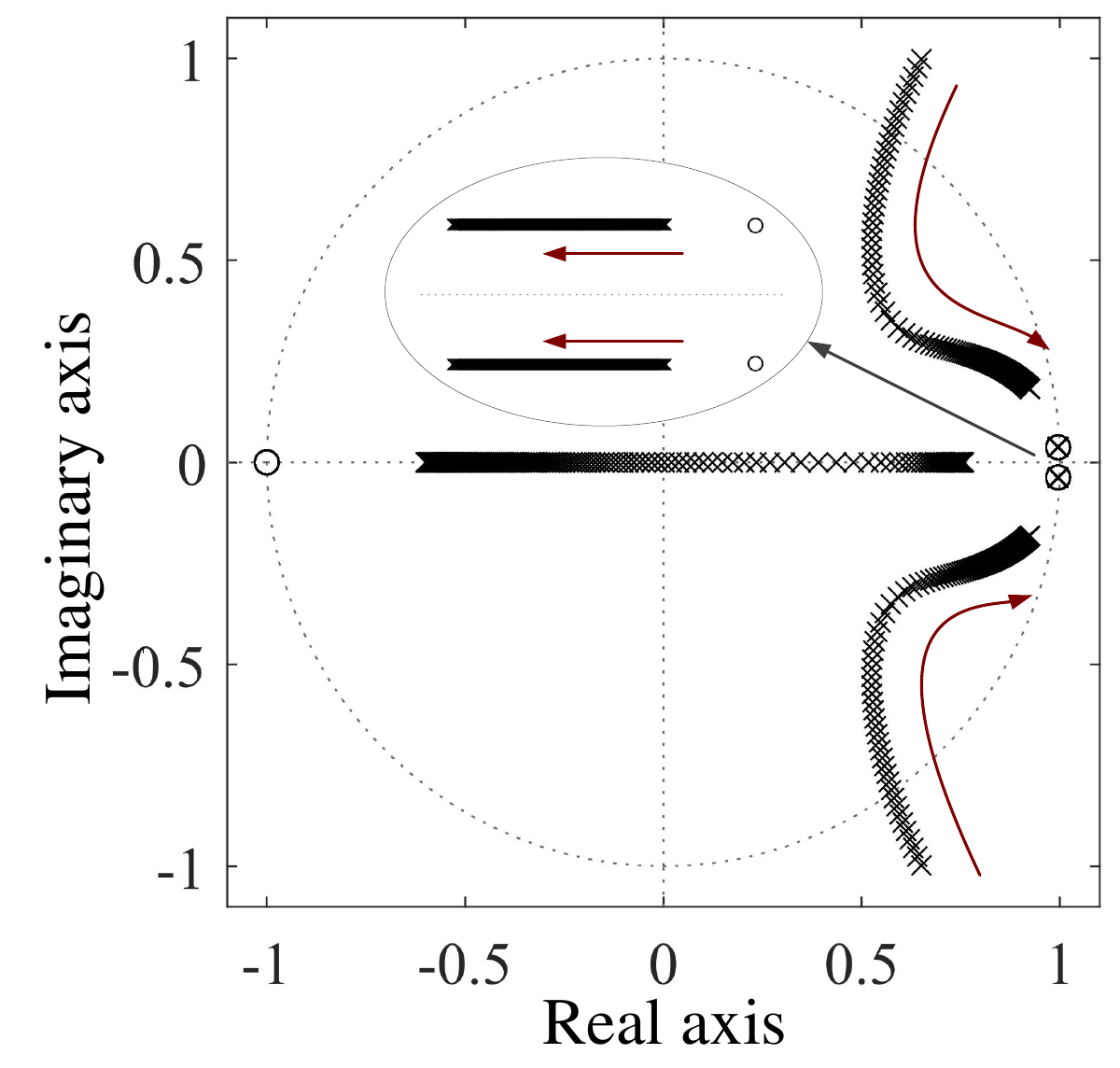}}

\vspace*{-0.4 cm}
\caption{Pole-zero map of the control loop with CVF-AD gain variation.}
\vspace*{-0.2 cm}

\label{fig:pole_zero_map}
\end{figure}

By incorporating an active damping loop into the controller and selecting an appropriate value of $K_a$, the converter stability is guaranteed. $K_a$ must be greater than the minimum value to damp the peak value of the resonance. For the study in this paper, the minimum value is calculated at around $3$ according to the LCL parameters. The damping increases by choosing a greater value, and the resonance can be damped sufficiently. As the grid inductance varies, $K_a$ is chosen to $12$ to guarantee stability even with grid inductance variation. Increasing the value of $K_a$ increases the damping ratio of the LCL filter; however, by increasing its value further, the LCL experiences over-damping, which may reduce the bandwidth and the fundamental harmonic. Fig.~\ref{fig:pole_zero_map} demonstrates the pole-zero map of the control loop with the variation of $K_a$. As the result shows, the system is unstable when $K_a$ is small, and when it increases, the system becomes stable. Hence, selecting the proper damping gain is necessary to avoid over-damping.

\section{Simulation Results}
To assess the proposed method's performance, simulation studies are conducted in MATLAB/Simulink. These studies examine a three-phase grid-connected inverter that is connected to the grid via an LCL filter. Table \ref{table:system_parameters} contains a list of rating voltages, currents, controller settings, and other AD specifications.
The sampling time is calculated based on the inverter's switching frequency, which is $10$~kHz in this study. According to the LCL filter design, the resonance frequency computes around $2.8$~kHz. In the simulation studies, the grid inductance has varying values ranging from $0.5$~mH to $6$~mH. Thus, the resonance range can vary between $1.35$~kHz and $2.1$~kHz based on the grid inductance. The maximum value of the grid impedance indicates a weak grid, which can occur when the transmission line length is too long. This condition is considered the worst-case scenario, and the minimum grid inductance represents a stiff grid, which is more stable and resilient to disturbances.

\begin{table}[t]
\caption{Grid and inverter parameters values.}

\vspace*{-0.6 cm}

\begin{center}
\begin{tabular}{cccccc}

\toprule
\textbf{Parameters} & \textbf{Symbol} & \textbf{Value}  & \textbf{Parameters}  & \textbf{Symbol} & \textbf{Value} \\ 
\toprule

\begin{tabular}[c]{@{}l@{}}Grid \\ voltage\end{tabular}         & $V_g$              & \begin{tabular}[c]{@{}l@{}}110 V\\ (RMS)\end{tabular} & \begin{tabular}[c]{@{}l@{}}Grid-side\\ inductor\end{tabular}    & $L_2$              & 0.4 mH         \\ 
\midrule
\begin{tabular}[c]{@{}l@{}}DC link \\ voltage\end{tabular}      & $V_{DC}$             & 350 V                                                 & \begin{tabular}[c]{@{}l@{}}Inverter-side\\ current\end{tabular} & $L_1$              & 1.6 mH         \\ 
\midrule
\begin{tabular}[c]{@{}l@{}}Fundamental\\ frequency\end{tabular} & $f_o$              & 60 Hz                                                 & \begin{tabular}[c]{@{}l@{}}LCL \\ capacitor\end{tabular}        & $C_f$              & 9.8 uF         \\ 
\midrule
\begin{tabular}[c]{@{}l@{}}Switching\\ frequency\end{tabular}   & $f_{sw}$             & 10 kHz                                                & \begin{tabular}[c]{@{}l@{}}Resonance\\ frequency\end{tabular}   & $F_r$              & 1.6 kHz        \\ 
\midrule
\begin{tabular}[c]{@{}l@{}}Proportional\\ gain\end{tabular}   &  $K_p$             &  8.3                                                & \begin{tabular}[c]{@{}l@{}}Resonance\\ gain \end{tabular}   & $K_r$              & 400         \\ 
\midrule

\begin{tabular}[c]{@{}l@{}}  AD  gain   \end{tabular}   & $K_a$   & 12   & \begin{tabular}[c]{@{}l@{}} Lead \\ pole \end{tabular}   &  $P_z$   & 0.75     \\ 

\bottomrule
\end{tabular}
\label{table:system_parameters}
\end{center}
\end{table}

Two case studies are considered to evaluate the proposed method for active damping.

\subsection{Case I: CVF-AD Activates After a While}
\begin{figure}
\centering
\includegraphics[width= 0.95\columnwidth ]{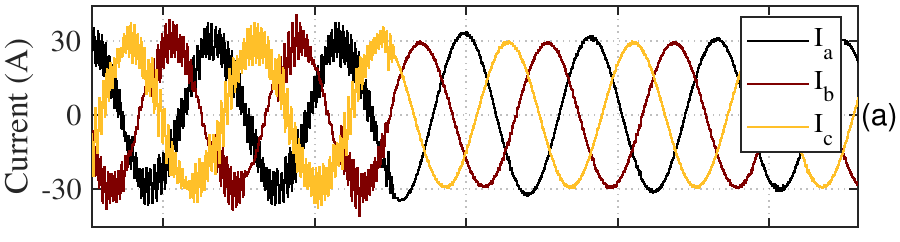}
\hfill
\includegraphics[width= 0.95\columnwidth ]{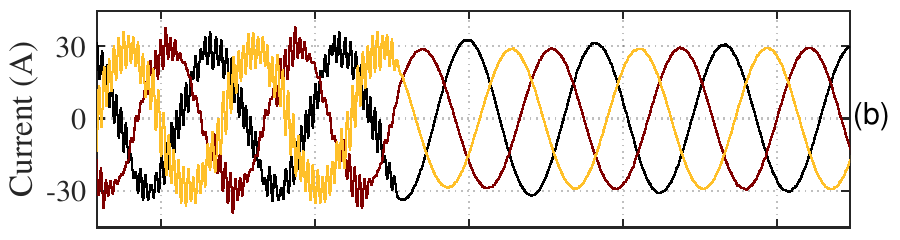}
\hfill
\includegraphics[width= 0.95\columnwidth ]{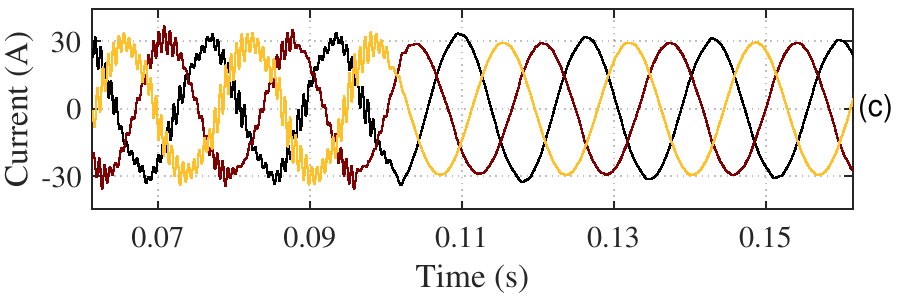}
\vspace*{-0.5 cm}
\caption{Grid-side current before and after CVF-AD activation. (a) For grid inductance  $L_g=0.5$~mH. (b) For $L_g=3$~mH. (c) For $L_g=6$~mH.}
\label{fig:3_phase_current_AD}
\end{figure}

In this case study, the inverter starts without AD, then, after a while, the inverter is equipped with AD to see the performance of the proposed method under various grid inductances. The system descriptions are shown in Table \ref{table:system_parameters}. The inverter starts without AD at $t=0$~ms.Then, at $t=100$~ms, the inverter control changes, and the inverter starts working with the CVF-AD with the proposed differentiator. As Fig.~\ref{fig:3_phase_current_AD} shows, the current is distorted due to the resonance in the system at first. The distorted current results from harmonic amplification at the resonance frequency and is added to the fundamental frequency. After activation of the CVF-AD, the output currents stop oscillations rapidly, and after a few cycles, the three-phase current stabilizes. Fig.~\ref{fig:3_phase_current_AD} shows the output current for various cases: $L_g=0.5$~mH, $L_g=3$~mH, $L_g=6$~mH. As a result, the active damping with the proposed differentiator can quickly stabilize the system and remove high-frequency harmonic in the system.

\subsection{Case II: An Increased Step Change in the Current}

\begin{figure}
\centering
\includegraphics[width= 0.95\columnwidth ]{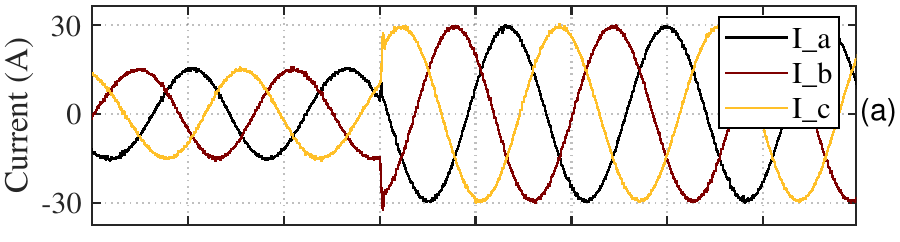}
\hfill
\includegraphics[width= 0.95\columnwidth ]{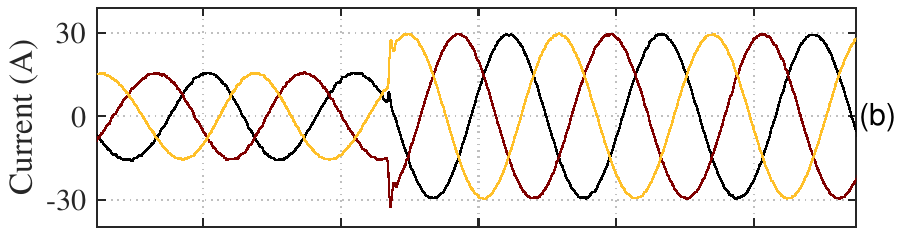}
\hfill
\includegraphics[width= 0.95\columnwidth ]{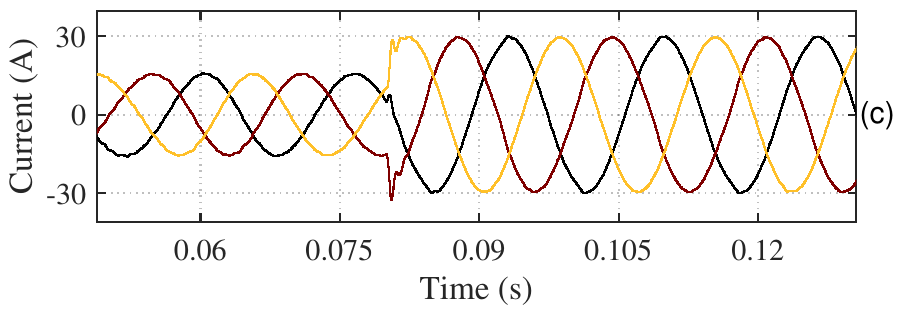}
\vspace*{-0.5 cm}
\caption{Step increase in the grid-side current. (a) For grid inductance, $L_g=0.5$~mH. (b) For $L_g=3$~mH. (c) For $L_g=6$~mH.}
\vspace*{-0.5 cm}
\label{fig:step_change_current}
\end{figure}

In the second case study, a three-phase inverter, same as in case I, starts with a CVF-AD that uses the proposed differentiator. In this case study, the current reference undergoes a step shift to raise the current under various grid inductances. The converter starts with the CVF-AD at $t=0$~s with a reference current value of $20$~A. At $t=80$~ms, the reference current increases, and the current gets $30$~A. 
Fig.~\ref{fig:step_change_current} (a), (b), and (c) show the grid-side current when the grid inductance is $L_g=0.5$~mH, $L_g=3$~mH, and $L_g=6$~mH, respectively. $L_g=6$~mH shows a weak grid in the simulations. As this figure shows, the controller follows the reference quickly. Therefore, the result demonstrated that the proposed differentiator mimics an ideal differentiator in the desired area, such that the active loop feedback is equal to the one that the capacitor current is directly fed back for damping purposes.


\section{Conclusion}

The LCL inherent resonance peak deteriorates the power quality and raises stability issues due to high-frequency amplification. This paper presents a novel method of differentiating the capacitor voltage feedback to eliminate the resonance in the grid-connected inverter. However, implementing the derivative term in practice gives rise to high-frequency noise and gets into the system, which is undesirable. To address this issue, this paper proposes a noise-immune function to replace the actual derivative term.
The performance of the proposed system is evaluated under grid impedance variation scenarios. Based on the simulation results, the CVF-AD effectively mitigates the resonance peak using the proposed differentiator.

\bibliographystyle{IEEEtran}
\bibliography{ref.bib}


\end{document}